\documentclass[showpacs,twocolumn,aps,prl]{revtex4}
\usepackage{amsmath}
\usepackage{amsfonts}
\usepackage{amssymb}
\usepackage{amsthm}
\usepackage{graphicx}
\usepackage{CJK}
\usepackage{color}

\begin{document}
\title{Statics and field-driven dynamics of transverse domain walls in biaxial nanowires under uniform transverse magnetic fields}
\author{Jie Lu}
\email{jlu@mail.hebtu.edu.cn}
\affiliation{College of Physics, Hebei Advanced Thin Films Laboratory, Hebei Normal University, Shijiazhuang 050024, China}
\date{\today}

\begin{abstract}
In this work, we report analytical results on transverse domain wall (TDW) statics and field-driven dynamics in quasi one-dimensional biaxial
nanowires under arbitrary uniform transverse magnetic fields (TMFs) based on the Landau-Lifshitz-Gilbert equation.
Without axial driving fields, the static TDW should be symmetric about its center meanwhile twisted in its azimuthal angle distribution.
By decoupling of polar and azimuthal degrees of freedom, an approximate solution is provided which reproduces these features to a great extent.
When an axial driving field is applied, the dynamical behavior of a TDW is viewed as the response of its static profile to external excitations.
By means of the asymptotic expansion method, the TDW velocity in traveling-wave mode is obtained, which provides the extent and boundary of the ``velocity-enhancement" effect of TMFs to TDWs in biaxial nanowires.
Finally numerical simulations are performed and strongly support our analytics.
\end{abstract}



\maketitle


\section{I. Introduction}
Domain wall (DW) dynamics in magnetic nanowires (NWs) has attracted much attention in the past few decades
because of not only academic interests but also application prospects in modern information industry.
Magnetic DWs can be basic bit units in modern logical\cite{Science_309_1688_2005} and storage\cite{Science_320_190_2008,Science_320_209_2008} devices.
By applying various driving factors,
such as magnetic fields\cite{Walker,Science_284_468_1999,nmat_2_85_2003,nmat_4_741_2005,EPL_78_57007_2007,PRL_100_127204_2008,PRL_104_037206_2010,
AGoussev_PRL,PRB_77_014413_2008,PRB_80_214426_2009,xrwfield_AOP,xrwfield_EPL},
spin-polarized currents\cite{Berger_PRB_1996,Slonczewski_JMMM_1996,PRL_92_077205_2004,PRL_92_207203_2004,PRL_96_197207_2006,PRL_97_057203_2006,
Science_330_1810_2010,PRL_105_157201_2010,APL_96_162506_2010},
or temperature gradients\cite{PRL_113_097201_2014,PRB_90_014414_2014,PRB_92_020410R_2015,PRB_92_064405_2015,PRB_92_140405_2015},
etc, DWs can be driven to
move along wire axis with quite high velocity, which results in high processing speed of devices
based on them. Traditionally, the time evolution of the magnetization distribution is
described by the nonlinear Landau-Lifshitz-Gilbert (LLG) equation\cite{LLG_equation}. Different driving factors manifest themselves
as different torque terms therein. For magnetic fields, it is the damping torque
that drives the DWs. For spin-polarized currents, the spin transfer torque (STT) from the conduction electrons
plays the role of the driver. While for temperature gradients, the driving force is the entropy or the magnonic STT.
Compared with the latter two, the magnetic-field-driven case is the oldest, but still quite active one.

Besides processing speed, another main concern in applications is the device integration level.
Advances in manufacturing thinner NWs greatly improve the device density meanwhile makes them
quasi one-dimensional (1D) systems. In these thin NWs,
head-to-head or tail-to-tail transverse DWs (TDWs) drift along the wire axis\cite{IEEE_Trans_Magn_33_4167_1997,JMMM_290_750_2005,PRB_76_184408_2007,Klaui_1}.
Their dynamics under axial driving fields can be understood by Walker's famous work in 1974\cite{Walker}.
According to his theory, the transverse magnetic anisotropy (TMA) plays a crucial role and leads to a critical
axial field strength named ``Walker limit" that separates two distinct propagation modes. Below it, TDWs undergo
a traveling-wave motion with a linear dependence of velocity on axial field strength.
While above it, the traveling-wave mode collapses (known as ``Walker breakdown") and TDWs
take a reciprocating rotation in which the drifting velocity dramatically decreases.
In recent works, the effects of spin waves in 1D magnetic NWs have been extensively studied\cite{PRL_98_087205_2007,PRL_107_177207_2011,PRL_109_167209_2012,
PRL_111_027205_2013,PRB_90_184415_2014,PRB_92_014411_2015}.
In reciprocal mode, the breathing effect of TDWs' width naturally generates spin waves\cite{PRL_109_167209_2012}.
What's impressive is that even in traveling-wave mode, spin waves are also emitted and make
Walker profile unstable\cite{PRL_111_027205_2013}. However, due to the finite damping, spin waves in real materials
will not survive too far from TDWs, thus saves the Walker solution.

To improve the processing speed, it is straightforward to suppress or at lease postpone the
occurrence of Walker breakdown. In recent years, several strategies have been proposed,
such as edge roughness\cite{nmat_2_521_2003}, additional underlayers with strong crystalline TMA\cite{APL_91_122513_2007}
or extra transverse magnetic fields (TMFs)\cite{JAP_103_073906_2008}, etc. Among them, using a TMF is the easiest way.
Thus it is of great value to systematically investigate the TDW dynamics under TMFs,
which is also an issue of interest in the academic community in recent years\cite{JAP_103_073906_2008,Kunz1,Kunz2,jlu_TMF_JAP,JAP_108_063904_2010,
Glathe1,Glathe2,Glathe3,JAP_106_113914_2009,APL_96_182507_2010,JKPS_2013,Klaui_2,JMMM_397_325_2016,
Sobolev1,Sobolev2,Sobolev3,arxiv1,arxiv2,AGoussev_PRB,AGoussev_Royal}.
Most existing works are numerical\cite{JAP_103_073906_2008,Kunz1,Kunz2,jlu_TMF_JAP,JAP_108_063904_2010}
or experimental\cite{Glathe1,Glathe2,Glathe3,JAP_106_113914_2009,APL_96_182507_2010,JKPS_2013,Klaui_2,JMMM_397_325_2016}.
In magnetic materials, the TMA from internal magnetic energy densities (crystalline, magnetostatic, etc)
are quadratic in magnetization while that from external magnetic energy density (Zeeman) is linear.
It is this mismatch in symmetry that increases the difficulty of theoretical analysis of TDW dynamics under TMFs.
To our knowledge, two series of theoretical efforts have been performed:
(1) In 1990s, Sobolev \textit{et al.} simplified the continuously twisted azimuthal angle distribution to
a plateau in a finite region with length $\pi\Delta$\cite{Sobolev1,Sobolev2,Sobolev3}.
Then by integrating the LLG equation over the entire wire, they obtained a set of generalized Slonczewski equations for TDW dynamics.
This approach has been widely used, including some recent works\cite{arxiv1,arxiv2}.
(2) In 2013, Goussev \textit{et al.} raised an asymptotic approach\cite{AGoussev_PRB,AGoussev_Royal}
to systematically explore TDW dynamics under TMFs. In this approach, the differential form of
LLG equation is preserved and expanded into series of some scaling parameter,
but the twisting of TDW in biaxial case is not fully considered.
In both strategies, the continuous form of twisted static TDW profile, which is an important issue in nanomagnetism, is absent.
Moveover, the extent and boundary of the ``velocity-enhancement" effect of TMFs to TDWs in biaxial NWs are unsolved.

This paper is organized as follows. In Sec. II we briefly introduce our 1D magnetic biaxial
system as well as the LLG equation. In Sec. III, by decoupling of the polar and azimuthal degrees of freedom,
the approximate static TDW profile under an arbitrary uniform TMF is presented.
In Sec. IV, the field-driven dynamics of TDWs under uniform
TMFs are investigated using the asymptotic approach,
hence provides the extent and boundary of the ``velocity-enhancement" effect of TMFs.
In Sec. V, OOMMF\cite{OOMMF} simulations are performed.
Finally, in Sec. VI we discuss the advantages and disadvantages of our approach.

\section{II. Modeling}
To begin with, a head-to-head TDW is nucleated in a long magnetic NW with thickness $t$ and
width $w$, as shown in Fig. 1. The $z$-axis is along wire axis, and $x$-axis is in the
thickness direction. $\Delta$ is the TDW width.
By assuming constant magnitude $M_s$, the magnetization
vector $\vec{M}(\vec{r})$ is fully described by its polar angle $\theta(\vec{r})$ and azimuthal angle $\phi(\vec{r})$.
For thin enough NWs, the magnetization in the wire cross-section should be uniform,
i.e. $\theta(\vec{r})\equiv\theta(z)$ and $\phi(\vec{r})\equiv\phi(z)$.
Finally, a uniform external TMF with magnitude $H_{\perp}$ and orientation angle $\Phi_{\perp}$,
\begin{equation}\label{TMF_vec}
    \vec{H}_{\mathrm{TMF}}=(H_x,H_y,0)=H_{\perp}(\cos\Phi_{\perp},\sin\Phi_{\perp},0)
\end{equation}
is exerted onto the entire wire and leads to a Zeeman energy density $-\mu_{0}\vec{M}\cdot\vec{H}_{\mathrm{TMF}}$.

\begin{figure}[htbp]
\centering
\scalebox{0.44}[0.44]{\includegraphics[angle=0]{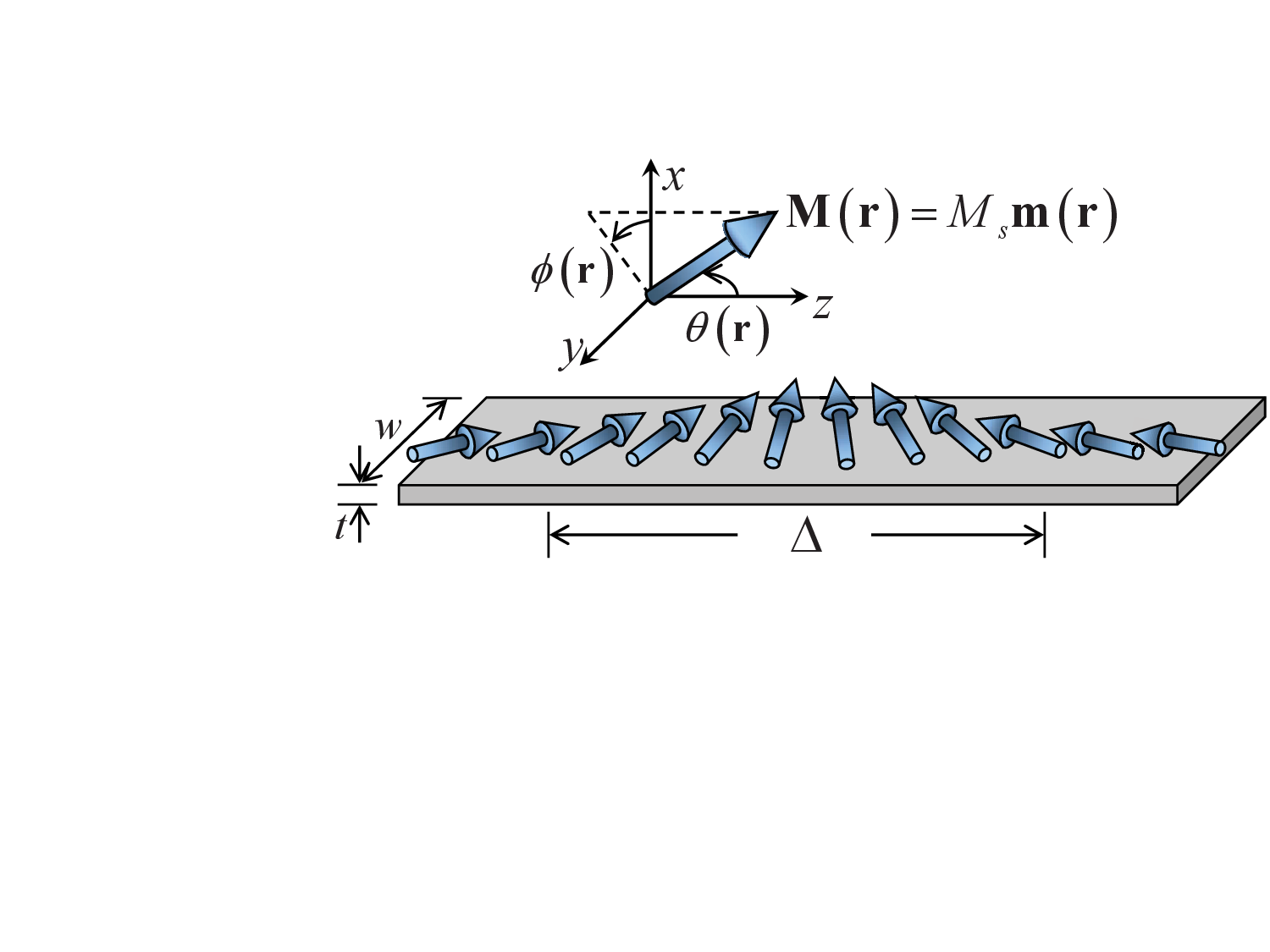}}
\caption{(Color online) Sketch of a head-to-head TDW in a nanowire with thickness $t$ and width $w$.
The coordinate system is as follows: $z-$axis is along wire axis, $x-$axis is in
thickness direction while $y-$axis is along $\hat{z}\times\hat{x}$. The TDW has a width $\Delta$.
$\theta(\vec{r})$ and $\phi(\vec{r})$ are the polar and azimuthal angles of the magnetization $\vec{M}(\vec{r})$, respectively.}\label{fig1}
\end{figure}

The time evolution of the magnetization $\vec{M}(\vec{r})\equiv M_{s}\vec{m}(\vec{r})$ is described
by the nonlinear LLG equation,
\begin{equation}\label{LLG_vec}
\frac{\partial\vec{m}}{\partial t}=-\gamma\vec{m}\times\vec{H}_
{\mathrm{eff}}+\alpha\vec{m}\times\frac{\partial\vec{m}}{\partial t}
\end{equation}
where $\gamma$ is the gyromagnetic ratio, $\vec{H}_{\mathrm{eff}}$ is the effective field and defined as
$-(1/\mu_{0}M_{s})(\partial E_{\mathrm{tot}}/\partial \vec{m})$ with $E_{\mathrm{tot}}$
being the total magnetic energy density. For our NW,
\begin{eqnarray}\label{E_density}
E_{\mathrm{tot}} &=& \frac{J}{M_s^2}(\nabla\vec{M})^2-\frac{1}{2}k_1^0 \mu_0 M_z^2+\frac{1}{2}k_2^0 \mu_0 M_x^2 \nonumber \\
   & & +E_{\mathrm{m}}(\vec{M})-\mu_0\vec{M}\cdot(\vec{H}_{\parallel}+\vec{H}_{\mathrm{TMF}})
\end{eqnarray}
where $\vec{H}_{\parallel}=H_1 \hat{e}_z$ is the external axial field,
$J$ is the exchange coefficient, and $k_1^0$($k_2^0$) is the crystalline anisotropy constant along the easy(hard) axis.
$E_{\mathrm{m}}$ denotes the magnetostatic energy density and is generally hard to be analyzed theoretically due to its non-local nature.
Our previous work\cite{xrwfield_EPL} showed that in thin enough NWs with rectangular cross-section, most of $E_{\mathrm{m}}$ can be described by
local quadratic terms of $M_{x,y,z}$ in terms of three average demagnetization factors $D_{x,y,z}$\cite{Aharoni_Dxyz}.
Therefore we have $k_1^0\rightarrow k_1=k_1^0+(D_y-D_z)$ and $k_2^0\rightarrow k_2=k_2^0+(D_x-D_y)$.

In spherical coordinates, the vectorial LLG equation (\ref{LLG_vec}) changes to scalar form,
\begin{subequations}\label{LLG_scalar_exact}
\begin{align}
  \gamma^{-1}\dot{\theta}(1+\alpha^2) = A-\alpha B   \\
  \gamma^{-1}\dot{\phi}(1+\alpha^2)\sin\theta =  B+\alpha A
\end{align}
\end{subequations}
with
\begin{subequations}\label{AB_exact}
\begin{align}
  A &\equiv  k_2 M_s \sin\theta\sin\phi\cos\phi -H_{\perp}\sin(\phi-\Phi_{\perp})   \nonumber \\
     & \quad +\frac{2J}{\mu_0 M_s\sin\theta}(\phi'\sin^2\theta)'   \\
  B &\equiv  M_s\sin\theta\cos\theta(k_1+k_2\cos^2\phi) - H_{\perp}\cos\theta\cos(\phi-\Phi_{\perp})  \nonumber \\
     & \quad  - \frac{2J}{\mu_0 M_s}\left[ \theta''-(\phi')^2\sin\theta\cos\theta \right] + H_1\sin\theta
\end{align}
\end{subequations}
where a dot (prime) means time (spatial) derivative.

When TMFs are absent, the Walker limit is $H_W=\alpha k_2 M_s/2$.
Below it, a TDW propagates as a traveling-wave with its azimuthal angle distribution being a plane,
\begin{equation}\label{Walker_phi_belowHW}
    \phi(z,H_1)=(1/2)\sin^{-1}(-H_1/H_W)
\end{equation}
and its velocity,
\begin{eqnarray}\label{Walker_velocity_belowHW}
 V_{\mathrm{Walker}} &=& \frac{\gamma\Delta_{\mathrm{Walker}}(H_1)}{\alpha} H_1   \nonumber \\
 \Delta_{\mathrm{Walker}}(H_1) &=& \Delta_0\left[ 1+(k_2/k_1)\cos^2\phi \right]^{-1/2}    \nonumber \\
 \Delta_0 &\equiv& 1/c = \sqrt{2J/(k_1\mu_0 M_s^2)}
\end{eqnarray}
Above it, the TMA torque can not balance the procession torque, hence the TDW undergoes a reciprocating rotation
with its drifting velocity greatly reduced, which is usually not preferred.

The above equations are all what we need when we start our work.

\section{III. Static profile of TDW in biaxial NWs under arbitrary uniform TMFs}
\subsection{A. Magnetization orientation in two domains}
To explore the static TDW profile, the first step is to calculate the magnetization orientations in the two faraway domains, namely the boundary condition.
In the left domain at $z\rightarrow-\infty$, the polar and azimuthal angles of magnetization are denoted as $\theta_{\infty}$
and $\phi_{\infty}$, respectively. While in the right domain at $z\rightarrow+\infty$, the two angles are $\pi-\theta_{\infty}$
and $\phi_{\infty}$, due to the symmetry about the wall center. The static condition,
\begin{equation}\label{static_condition}
    \dot{\theta}_{\infty}=0,\quad \dot{\phi}_{\infty}=0
\end{equation}
and the domain condition,
\begin{equation}\label{domain_condtion}
    \theta'_{\infty}=\theta''_{\infty}=0,\quad \phi'_{\infty}=\phi''_{\infty}=0
\end{equation}
as well as the existence condition of TDW,
\begin{equation}\label{TDW_existence_condition}
   \theta_{\infty}\neq \pi/2
\end{equation}
make the LLG equation (\ref{LLG_scalar_exact}) reduced to,
\begin{subequations}\label{AB_static_in_domains}
\begin{align}
 H_{\perp}\sin(\phi_{\infty}-\Phi_{\perp}) &= k_2 M_s \sin\theta_{\infty}\sin\phi_{\infty}\cos\phi_{\infty}  \\
H_{\perp}\cos(\phi_{\infty}-\Phi_{\perp}) &= M_s\sin\theta_{\infty}(k_1+k_2\cos^2\phi_{\infty})
\end{align}
\end{subequations}
The summation of the squares of Eqs. (\ref{AB_static_in_domains}a,b) gives the polar angle in the left domain, $\theta_L$, as
\begin{equation}\label{theta_in_domains}
    \theta_L=\theta_{\infty}=\sin^{-1}\left(H_{\perp}/H_{\perp}^{\mathrm{max}}\right)
\end{equation}
where
\begin{equation}\label{H_perp_max_definition}
  H_{\perp}^{\mathrm{max}}=k_1 M_s\left[1+\frac{k_2(2k_1+k_2)}{k_1^2+(k_1+k_2)^2\tan^2\Phi_{\perp}}\right]^{\frac{1}{2}}
\end{equation}
and that in the right domain is $\theta_R=\pi-\theta_L$.
Eq. (\ref{theta_in_domains}) means a TMF must lift the magnetization in the two domains away from the wire axis.
Meanwhile its strength must be smaller than $H_{\perp}^{\mathrm{max}}$, otherwise the TDW structure can not survive.
The ratio of Eqs. (\ref{AB_static_in_domains}a,b) gives the azimuthal angle in the two domains,
\begin{equation}\label{phi_in_domains}
    \phi_{\infty}=\tan^{-1}\left(\frac{k_1+k_2}{k_1}\tan\Phi_{\perp}\right)
\end{equation}
This means the $\phi$-plane in the two domains always lies between the TMF and the easy $yz$-planes.
Note that ``$\phi_{\infty}+\pi$" is also a solution, but not stable since it has higher Zeeman energy.

\subsection{B. Twisting of TDW in $\phi$}
A static TDW under a general (not in easy or hard planes) uniform TMF must have twisting in its $\phi$-plane.
To see this, suppose there is no twisting at all in the entire TDW region.
This means: (a) $\phi(z)$ keeps the same value $\phi_{\infty}$ as those in the two domains, and thus satisfies $\phi'=\phi''\equiv 0$;
(b) $\theta(z)$ varies continuously from $\theta_L$ to $\theta_R$.
Hence the static condition, $A=B=0$, is reduced to,
\begin{subequations}\label{AB_static_in_TDW}
\begin{align}
  H_{\perp}\sin(\phi-\Phi_{\perp}) &=  k_2 M_s \sin\theta\sin\phi\cos\phi  \\
  H_{\perp}\cos\theta\cos(\phi-\Phi_{\perp}) &=  M_s\sin\theta\cos\theta(k_1+k_2\cos^2\phi) \nonumber \\
                                                     & \quad -\frac{2J}{\mu_0 M_s}\theta''
\end{align}
\end{subequations}
From Eqs. (\ref{AB_static_in_domains}a) and (\ref{phi_in_domains}), Eq.(\ref{AB_static_in_TDW}a) always holds
when $\Phi_{\perp}=n\pi/2$ with $n$ being an integer ($\phi\equiv\phi_{\infty}=\Phi_{\perp}$, $\theta$ can be arbitrary).
In this situation, the solution of Eq. (\ref{AB_static_in_TDW}b) gives the $\theta$-profile of the TDW as,
\begin{eqnarray}\label{theta_static_in_TDW}
  \tan\frac{\theta}{2} &=& \frac{e^{z_1}+\tan\frac{\theta_{\infty}}{2}}{1+e^{z_1}\tan\frac{\theta_{\infty}}{2}}  \nonumber \\
  z_1 &=& z\cos\theta_{\infty}/\Delta(\phi_{\infty})   \nonumber \\
  \Delta(\phi_{\infty}) &=& \Delta_0 \left[1+(k_2/k_1)\cos^2\phi_{\infty}\right]^{-1/2}
\end{eqnarray}
where $\theta_{\infty}$ and $\phi_{\infty}$ are given in Eqs. (\ref{theta_in_domains}) and (\ref{phi_in_domains}), respectively.
This implies that TDWs in easy planes ($n=2m+1$) will be wider than those in hard ones ($n=2m$).

However, if $\Phi_{\perp}\neq n \pi /2$, $\phi_{\infty}$ is no longer equal to $\Phi_{\perp}$ (see Eq.(\ref{phi_in_domains}))
and then Eq. (\ref{AB_static_in_TDW}a) can not always hold.
This contradiction comes entirely from the hypothesis of $\phi(z)\equiv\phi_{\infty}$.
Hence, a twisting of $\phi$-plane must appear in the TDW region, which explains the existing experimental observations\cite{Klaui_1}.

\subsection{C. Decoupling of $\theta$ and $\phi$}

Rewrite the general static condition, $A=B=0$, as
\begin{subequations}\label{AB_to_f1f2}
\begin{align}
  0 &= f_1(\theta,\phi)  \\
  \frac{2J}{\mu_0 M_s}\theta'' &= M_s(k_1+k_2\cos^2\phi_{\infty})(\sin\theta-\sin\theta_{\infty})\cos\theta \nonumber \\
    & \quad + f_2(\theta,\phi)\cos\theta
\end{align}
\end{subequations}
with
\begin{subequations}\label{f1f2_definition}
\begin{align}
  f_1(\theta,\phi) &\equiv -H_{\perp}\sin(\phi-\Phi_{\perp})+k_2M_s\sin\theta\sin\phi\cos\phi    \nonumber \\
   & \quad +\frac{2J}{\mu_0 M_s}(2\theta'\phi'\cos\theta+\phi''\sin\theta)  \\
 f_2(\theta,\phi) &\equiv \frac{2J}{\mu_0 M_s}(\phi')^2\sin\theta+k_2 M_s\sin\theta(\cos^2\phi-\cos^2\phi_{\infty})   \nonumber \\
   & \quad +H_{\perp}\left[\cos(\phi_{\infty}-\Phi_{\perp})-\cos(\phi-\Phi_{\perp})\right]
\end{align}
\end{subequations}

The nonlinearity of LLG equation makes $\theta$ and $\phi$ coupled with each other in Eq. (\ref{AB_to_f1f2}).
Now we try to decouple them meanwhile preserve the continuous twisting in $\phi$ as much as possible.
To do this, in Eq.(\ref{AB_to_f1f2}b) we drop the last term, which is equivalent to directly let
\begin{equation}\label{theta_static trial_equation}
    \frac{2J}{\mu_0 M_s}\theta'' = M_s(k_1+k_2\cos^2\phi_{\infty})(\sin\theta-\sin\theta_{\infty})\cos\theta
\end{equation}
The solution of Eq. (\ref{theta_static trial_equation}) has been provided in Eq. (\ref{theta_static_in_TDW}), which means that
the polar angle $\theta(\theta_{\infty}<\theta<\pi-\theta_{\infty})$ still takes the ``Walker-ansatz" form.

Next we turn to $\phi$-profile. To realize the above decoupled $\theta$-profile, we must have,
\begin{equation}\label{f1f2_eq_0}
    f_1(\theta,\phi)=f_2(\theta,\phi)=0
\end{equation}
By noting that $\partial\phi/\partial z$ is not always zero (i.e. twisting), from Eq. (\ref{f1f2_eq_0}) one has,
\begin{equation}\label{f1f2_weakened}
    \phi'\cdot f_1+\frac{\partial f_2}{\partial z}=0
\end{equation}
This leads to,
\begin{equation}\label{phi_static_trial_equation_original}
    \left[\frac{6J}{k_2\mu_0 M_s^2}(\phi')^2+(\cos^2\phi-\cos^2\phi_{\infty})\right]\sin^2\theta\equiv \mathrm{const}
\end{equation}
Considering the fact that at $z=\pm\infty$, $\phi\equiv\phi_{\infty}$ and $\phi'\equiv 0$, we have $\mathrm{const}\equiv 0$.
This gives us the differential equation that $\phi(z)$ should satisfy inside the TDW,
\begin{equation}\label{phi_static_trial_equation}
    (\phi')^2=\frac{\sin^2\phi-\sin^2\phi_{\infty}}{(\Delta_2)^2},\quad \Delta_2=\sqrt{\frac{6J}{k_2\mu_0 M_s^2}}
\end{equation}

Before solving this equation, we point out an important feature of its resulting $\phi$-profile: \emph{it has a finite twisting region}.
To see this, by differentiating Eq. (\ref{phi_static_trial_equation}) with respect to $z$, we have,
\begin{equation}\label{finite_twisting_region_discussion}
    \phi'\phi''=\frac{\sin\phi\cos\phi}{(\Delta_2)^2}\phi'
\end{equation}
Take $0<\Phi_{\perp}<\pi/2$ as an example. From Eq. (\ref{phi_in_domains}), $0<\Phi_{\perp}<\phi_{\infty}<\pi/2$.
If $\phi'$ does not vanish at $z=\pm\infty$, we must have $\phi''>0$, which is contradictory with $\phi(z=\pm\infty)\equiv\phi_{\infty}$.
This implies that the twisting must vanish at some
finite coordinate, say $\pm z^*$ (``$\pm$" comes from the symmetry about the wall center). At these two points, $\phi'$ is zero and continuous,
while $\phi''$ may be discontinuous.

Generally, Eq. (\ref{phi_static_trial_equation}) has multiple solutions, depending on which azimuthal angle ($\phi_{\infty}$ or $\phi_{\infty}+\pi$)
is selected in each domain. For a stable solution, $\phi_{\infty}$ is selected in both domains, which means
\begin{equation}\label{phi_static_BC}
    \phi_{z=\pm\infty}=\phi_{\infty}=\tan^{-1}\left(\frac{k_1+k_2}{k_1}\tan\Phi_{\perp}\right)
\end{equation}
When $\Phi_{\perp}=n\pi/2$, there is no twisting and Eq. (\ref{phi_static_trial_equation}) has a trivial solution:
\begin{equation}\label{phi_static_profile_trivial}
    \phi(z)\equiv\phi_{\infty}=\Phi_{\perp}
\end{equation}
When $\Phi_{\perp}\neq n\pi/2$, the solution becomes complicated. Without losing generality, we assume that $0<\Phi_{\perp}<\pi/2$.
Combining Eqs. (\ref{phi_static_trial_equation}) and (\ref{phi_static_BC}), we obtain the final solution as follows,
\begin{eqnarray}\label{phi_static_in_TDW}
 \frac{|z|-z_2}{\Delta_2} &=& F\left(\cos\phi_{\infty},\cos^{-1}\frac{\sqrt{\sin^2\phi-\sin^2\phi_{\infty}}}{\cos\phi_{\infty}}\right) \nonumber \\
 & & \cdot \mathrm{sgn}\left(\frac{\pi}{2}-\phi\right)  \nonumber \\
 z_2 &=& \Delta_2 F(\cos\phi_{\infty})  \nonumber \\
|z| &\leq& 2z_2,\quad \phi_{\infty}\leq \phi \leq \pi-\phi_{\infty}
\end{eqnarray}
where $F(k)=\int_0^{\pi/2}\frac{\mathrm{d}\omega}{\sqrt{1-k^2\sin^2\omega}},\:0<k<1$ is the complete elliptic integral of the first kind,
$F(k,\eta)=\int_0^{\eta}\frac{\mathrm{d}\omega}{\sqrt{1-k^2\sin^2\omega}},\:0<k<1,\:0\leq\eta\leq\pi/2$ is the incomplete elliptic
integral of the first kind, and $\mathrm{sgn}(x)$ is the sign function. The solution (\ref{phi_static_in_TDW}) has several interesting features:
(1) it is symmetric about the TDW center $z=0$;
(2) at $z=0$, the twisting of $\phi$-plane becomes maximum and equals to $\pi-2\phi_{\infty}$;
(3) at $z=\pm z_2$, $\phi(z)=\pi/2$;
(4) when $|z|<2z_2$, TDW is twisted; when $|z|\geq2z_2$, $\phi(z)\equiv \phi_{\infty}$;
(5) at $z=\pm z^*=\pm2 z_2$, $\phi$ and $\phi'$ are continuous but $\phi''$ is discontinuous (see Eq. (\ref{finite_twisting_region_discussion})).

In summary, Eqs. (\ref{theta_static_in_TDW}) and (\ref{phi_static_in_TDW}) provide the entire approximate solution of static TDW profile
in a biaxial NW under an arbitrary uniform TMF. Compared with the polar angle profile (\ref{theta_static_in_TDW}),
the azimuthal angle profile (\ref{phi_static_in_TDW}) is less accurate due to the extra
approximation in Eq. (\ref{f1f2_weakened}). The main fault of (\ref{phi_static_in_TDW}) is that it overestimates the twisting magnitude.
In reality (see simulation results in Sec. V), the azimuthal angle of the magnetization at the TDW center
lies in the same quadrant with those in the faraway domains and extremely approaches the easy plane to minimize the anisotropic energy.
Thus the real maximum twisting angle is around $\pi/2-\phi_{\infty}$, which is half of that from (\ref{phi_static_in_TDW}).
This implies that Eq. (\ref{phi_static_in_TDW}) will get worse when TMFs approach the hard axis.

At last, our solution will reduce to a newly derived analytical static TDW profile
provided in Ref. \cite{JMMM_2015_Hertel} in the absence of TMFs as long as $k_1\rightarrow N_{\phi}$ and $k_2\rightarrow 0$.
This is because in that work there is no crystalline anisotropy.
The magnetostatic interaction induces an effective easy axis lying along the wire axis,
and no hard axis appears due to the circular symmetry in the cross-section.
In brief, our static solution can be viewed as a generalization of the one in Ref. \cite{JMMM_2015_Hertel}
to biaxial case when a hard axis and a uniform TMF coexist.

\section{IV. Field-driven TDW dynamics in biaxial NWs under TMFs}
The main idea of the asymptotic approach\cite{AGoussev_PRB,AGoussev_Royal} is to view the dynamical behavior of a TDW
 as a response of its static profile to external excitations (that is, axial driving fields), which leads to simultaneous
 scaling of field and velocity (or inverse of time).  Compared with uniaxial case, the twisting of
 azimuthal plane in biaxial case results in two problems: (a) zero-order solution becomes hard to solve; and (b) whether
 the self-adjoint Schr\"{o}dinger operator still holds.
 We here present our answer to these questions for both small and finite TMF cases.
 Since the asymptotic approach belongs to the linear response framework, we limit ourselves in the traveling-wave mode.
In particular, we will provide the extent and boundary of the ``velocity-enhancement" effect of TMFs in biaxial NWs.

\subsection{A. Small TMF case}
In this case, the fields and inverse of time are rescaled simultaneously as follows,
\begin{equation}\label{smallTMF_scaling_fields_and_time}
    H_{1,\perp} =\epsilon h_{1,\perp},\quad 1/t=\epsilon (1/\tau)
\end{equation}
where $\epsilon$ is a dimensionless infinitesimal.
We want to seek for a solution of the LLG equation with the following series expansion form:
\begin{subequations}\label{smallTMF_series_expansion_old}
\begin{align}
  \theta(z,t) &= \theta_0(z,\tau)+\epsilon\theta_1(z,\tau)+O(\epsilon^2) \\
  \phi(z,t) &= \phi_0(z,\tau)+\epsilon\phi_1(z,\tau)+O(\epsilon^2)
\end{align}
\end{subequations}
meanwhile obeying the boundary condition,
\begin{equation}\label{smallTMF_BC}
    \vec{M}(\pm\infty,\tau)=(\epsilon h_{\perp}\cos\Phi_{\perp},\epsilon h_{\perp}\sin\Phi_{\perp},\pm M_s)+O(\epsilon^2)
\end{equation}
Puttting Eq. (\ref{smallTMF_series_expansion_old}) into the LLG equation (\ref{LLG_scalar_exact}), to the zero order of $\epsilon$, we have,
\begin{subequations}\label{smallTMF_0order_equations}
\begin{align}
  0 &= (\phi'_0\sin^2\theta_0)'+c^2\frac{k_2}{k_1}\sin^2\theta_0\sin\phi_0\cos\phi_0     \\
  0 &= \theta''_0-\frac{1}{2} \left[c^2\left(1+\frac{k_2}{k_1}\cos^2\phi_0\right)+(\phi'_0)^2\right]\sin 2 \theta_0
\end{align}
\end{subequations}
This implies that the zero-order solution $\theta_0(z)$, $\phi_0(z)$
makes $\vec{H}_{\mathrm{eff}}(z)\| \vec{M}(z)$, which is just the static condition.
The stable solution of Eq. (\ref{smallTMF_0order_equations}) that satisfies Eq. (\ref{smallTMF_BC}) has the following form:
\begin{subequations}\label{smallTMF_0order_solution}
\begin{align}
  \phi_0(z,\tau) &\equiv (n+1/2)\pi  \\
  \theta_0(z,\tau) &= 2\tan^{-1} [e^{c(z-z_0(\tau))}]
\end{align}
\end{subequations}
which means the azimuthal plane is fixed in the easy $yz$-plane (due to the $k_2$-terms in Eq. (\ref{smallTMF_0order_equations})),
and the TDW center $z_0$ is a function of $\tau$ only.
By introducing a traveling coordinate $\xi\equiv z-z_0$, Eq. (\ref{smallTMF_series_expansion_old}) is rewritten as,
\begin{subequations}\label{smallTMF_series_expansion_new}
\begin{align}
  \theta(z,t) &= \theta_0(\xi,\tau)+\epsilon\theta_1(\xi,\tau)+O(\epsilon^2) \\
  \phi(z,t) &= \phi_0(\xi,\tau)+\epsilon\phi_1(\xi,\tau)+O(\epsilon^2)
\end{align}
\end{subequations}
To obtain the TDW velocity ``$\epsilon \mathrm{d}z_0/\mathrm{d}\tau$", we should proceed to the next order of $\epsilon$.
After some algebra, the first order equation can be finally
written as (from now on in this section, a ``prime" means partial derivative with respect to $\xi=z-z_0$),
\begin{eqnarray}\label{smallTMF_Ltheta1_f}
  \mathbf{L}\theta_1 &=& f      \nonumber \\
  \mathbf{L}&\equiv & \frac{2J}{\mu_0 M_s}\left(-\frac{\partial^2}{\partial \xi^2}+\frac{\theta'''_0}{\theta'_0}\right)  \nonumber \\
  f &\equiv &  \frac{c\alpha\sin\theta_0}{\gamma}\frac{\mathrm{d}z_0}{\mathrm{d}\tau}-h_1\sin\theta_0+h_{\perp}\cos\theta_0\sin\Phi_{\perp}
\end{eqnarray}
and
\begin{eqnarray}\label{smallTMF_Mphi1_g}
  \mathbf{M}\phi_1 &=& g      \nonumber \\
  \mathbf{M}&\equiv& \frac{2J}{\mu_0 M_s\sin\theta_0}\frac{\partial}{\partial \xi}\sin^2\theta_0\frac{\partial}{\partial \xi}-k_2 M_s \sin\theta_0  \nonumber \\
  g &\equiv&  \frac{-c\sin\theta_0}{\gamma}\frac{\mathrm{d}z_0}{\mathrm{d}\tau}+h_{\perp}\cos\Phi_{\perp}
\end{eqnarray}
Here $\mathbf{L}$ is the same 1D self-adjoint Schr\"{o}dinger operator as given in Refs. \cite{AGoussev_PRB,AGoussev_Royal}.
Following the ``Fredholm alternative",
by demanding $\theta'_0$ (i.e. the kernel of $\mathbf{L}$) to be orthogonal to the function $f$
defined in Eq. (\ref{smallTMF_Ltheta1_f}), and noting that $\langle\theta'_0,1\rangle=\pi$, $\langle\theta'_0,\sin\theta_0\rangle=2$
and $\langle\theta'_0,\cos\theta_0\rangle=0$, we obtain the TDW velocity in traveling-wave mode,
\begin{equation}\label{smallTMF_velocity}
    V_a=\epsilon\frac{\mathrm{d}z_0}{\mathrm{d}\tau}=\frac{\gamma\epsilon h_1}{c\alpha}=\frac{\gamma\Delta_0}{\alpha}H_1
\end{equation}

\subsection{B. Finite TMF case}
When the TMF strength $H_{\perp}$ becomes finite, we rescale the axial driving field $H_1$ and the TDW velocity $V_b$ simultaneously,
\begin{equation}\label{finiteTMF_scaling}
    H_1=\epsilon h_1,\quad V_b=\epsilon v_b
\end{equation}
By defining the traveling coordinate
\begin{equation}\label{finiteTMF_xi}
    \xi\equiv z-V_b t=z-\epsilon v_b t
\end{equation}
the traveling-wave solution $\theta(z,t)$, $\phi(z,t)$ are expanded as follows,
\begin{subequations}\label{finiteTMF_series_expansion}
\begin{align}
  \theta(z,t) &= \theta_0(\xi)+\epsilon\theta_1(\xi)+O(\epsilon^2) \\
  \phi(z,t) &= \phi_0(\xi)+\epsilon\phi_1(\xi)+O(\epsilon^2)
\end{align}
\end{subequations}
Substituting Eq. (\ref{finiteTMF_series_expansion}) into the LLG equation (\ref{LLG_scalar_exact}),
to the zero order of $\epsilon$, one has (from now on in this section, a ``prime" means partial derivative with respect to $\xi=z-V_b t$),
\begin{subequations}\label{finiteTMF_0order_equations_A0B0}
\begin{align}
  0 &=A_0=H_{\perp}\sin(\Phi_{\perp}-\phi_0)+k_2 M_s \sin\theta_0\sin\phi_0\cos\phi_0  \nonumber \\
    & \quad  +\frac{2J}{\mu_0 M_s}(2\theta'_0\phi'_0\cos\theta_0+\phi''_0\sin\theta_0)  \\
  0 &= B_0=-H_{\perp}\cos\theta_0\cos(\Phi_{\perp}-\phi_0)-\frac{2J}{\mu_0 M_s}\theta''_0   \nonumber \\
   & \quad +k_1 M_s\sin\theta_0\cos\theta_0\left[1+\frac{k_2}{k_1}\cos^2\phi_0+\frac{(\phi'_0)^2}{c^2}\right]
\end{align}
\end{subequations}
which are exactly the same with the static equations (\ref{AB_to_f1f2}-\ref{f1f2_definition}).
Therefore, the solution of the zero-order equations is just Eqs. (\ref{theta_static_in_TDW}) and (\ref{phi_static_in_TDW}),
with boundary conditions (\ref{theta_in_domains}) and (\ref{phi_in_domains}).
To obtain the TDW propagation velocity, we need to proceed to the next order.

At the first order of $\epsilon$, LLG equation becomes,
\begin{subequations}\label{finiteTMF_1order_equations}
\begin{align}
  (-v_b)\gamma^{-1}(1+\alpha^2)\theta'_0 &= A_1-\alpha B_1    \\
  (-v_b)\gamma^{-1}(1+\alpha^2)\sin\theta_0\phi'_0 &= B_1+\alpha A_1
\end{align}
\end{subequations}
or equivalently,
\begin{subequations}\label{finiteTMF_1order_equations_A1B1}
\begin{align}
  A_1 &= -v_b\gamma^{-1}(\theta'_0+\alpha\sin\theta_0 \phi'_0)   \\
  B_1 &= -v_b\gamma^{-1}(-\alpha\theta'_0+\sin\theta_0\phi'_0)
\end{align}
\end{subequations}
where
\begin{eqnarray}\label{finiteTMF_A1}
  A_1 &=& \mathbf{P}\theta_1+\mathbf{Q}\phi_1    \nonumber \\
  \mathbf{P} &\equiv& k_2 M_s\cos\theta_0\sin\phi_0\cos\phi_0     \nonumber \\
     & & +\frac{2J}{\mu_0 M_s}\left[2\phi'_0\left(\cos\theta_0\cdot\frac{\partial}{\partial\xi}-\theta'_0\sin\theta_0\right)+\phi''_0\cos\theta_0\right]   \nonumber \\
  \mathbf{Q} &\equiv& -H_{\perp}\cos(\Phi_{\perp}-\phi_0)+k_2 M_s \sin\theta_0\cos 2\phi_0   \nonumber \\
     & &  +\frac{2J}{\mu_0 M_s}\left(2\theta'_0\cos\theta_0\cdot\frac{\partial}{\partial\xi}+\sin\theta_0\cdot\frac{\partial^2}{\partial\xi^2}\right)
\end{eqnarray}
and
\begin{eqnarray}\label{finiteTMF_B1}
 B_1 &=&h_1\sin\theta_0+\mathbf{R}\theta_1+\mathbf{S}\phi_1  \nonumber \\
\mathbf{R} &\equiv& k_1 M_s\cos 2\theta_0\left[1+\frac{k_2}{k_1}\cos^2\phi_0+\frac{(\phi'_0)^2}{c^2}\right]  \nonumber \\
       & & +H_{\perp}\sin\theta_0\cos(\Phi_{\perp}-\phi_0)-\frac{2J}{\mu_0 M_s}\frac{\partial^2}{\partial\xi^2}    \nonumber \\
\mathbf{S} &\equiv& k_1 M_s\sin 2\theta_0\left(\frac{\phi'_0}{c^2}\frac{\partial}{\partial\xi}-\frac{k_2}{k_1}\sin\phi_0\cos\phi_0\right)   \nonumber \\
       & & -H_{\perp}\cos\theta_0\sin(\Phi_{\perp}-\phi_0)
\end{eqnarray}

We need to simplify the operators $\mathbf{R}$ and $\mathbf{S}$ to obtain the $v_b(h_1)$ relationship.
The zero-order equation (\ref{finiteTMF_0order_equations_A0B0}b) provides essential information.
First we assume that $\theta_0$ and $\phi_0$ have been decoupled. By partially differentiating ``$B_0=0$"
with respect to $\theta_0$, we have,
\begin{eqnarray}\label{finiteTMF_B0eq0_theta0}
   \frac{2J}{\mu_0 M_s}\frac{\theta'''_0}{\theta'_0} &=& k_1 M_s\cos 2\theta_0\left[1+\frac{k_2}{k_1}\cos^2\phi_0+\frac{(\phi'_0)^2}{c^2}\right]     \nonumber \\
   & &  +H_{\perp}\sin\theta_0\cos(\Phi_{\perp}-\phi_0)
\end{eqnarray}
This relationship simplifies the operator $\mathbf{R}$ to,
\begin{equation}\label{finiteTMF_R_simplify}
    \mathbf{R}=\frac{2J}{\mu_0 M_s}\left(-\frac{\partial^2}{\partial\xi^2}+\frac{\theta'''_0}{\theta'_0}\right)\equiv \mathbf{L}
\end{equation}
In addition, by partially differentiating $B_0=0$ with respect to $\phi_0$, we have,
\begin{eqnarray}\label{finiteTMF_B0eq0_phi0}
0 & = & k_1 M_s\sin 2\theta_0\left(\frac{\phi''_0}{c^2}-\frac{k_2}{k_1}\sin\phi_0\cos\phi_0\right)  \nonumber \\
   & & -H_{\perp}\cos\theta_0\sin(\Phi_{\perp}-\phi_0)
\end{eqnarray}
Eq. (\ref{finiteTMF_B0eq0_phi0}) simplifies the operator $\mathbf{S}$ to
\begin{equation}\label{finiteTMF_S_simplify}
\mathbf{S}=\frac{k_1 M_s \sin 2\theta_0}{c^2}\left(\phi'_0\frac{\partial}{\partial\xi}-\phi''_0\right)
\end{equation}
Remember in zero-order profile, twisting only occurs around
TDW center, where $\theta_0\sim\pi/2$ and $\sin 2\theta_0\approx 0$,
thus we can reasonably let $\mathbf{S}\approx 0$. Finally, the first-order equation (\ref{finiteTMF_1order_equations_A1B1}b) turns to the following form,
\begin{equation}\label{finiteTMF_Ltheta1}
    \mathbf{L}\theta_1=-h_1\sin\theta_0+(-v_b)\gamma^{-1}(-\alpha\theta'_0+\sin\theta_0\phi'_0)
\end{equation}
Again, the right hand side of Eq. (\ref{finiteTMF_Ltheta1}) must be orthogonal to the kernel of $\mathbf{L}$, i.e. $\theta'_0$,
in order that a solution exists. Noting that $\langle\theta'_0,1\rangle=\pi-2\theta_{\infty}$, $\langle\theta'_0,\sin\theta_0\rangle=2\cos\theta_{\infty}$,
$\langle\theta'_0,\theta'_0\rangle=[2\cos\theta_{\infty}-(\pi-2\theta_{\infty})\sin\theta_{\infty}]/\Delta(\phi_{\infty})$
and $\langle\theta'_0,\phi'_0\sin\theta_0\rangle=0$,
the TDW velocity finally reads,
\begin{equation}\label{finiteTMF_velocity}
    V_b=\frac{2\cos\theta_{\infty}}{2\cos\theta_{\infty}-(\pi-2\theta_{\infty})\sin\theta_{\infty}} \cdot \frac{\gamma\Delta(\phi_{\infty})}{\alpha}H_1
\end{equation}
in which $\theta_{\infty}$, $\phi_{\infty}$ and $\Delta(\phi_{\infty})$ are provided in Eqs. (\ref{theta_in_domains}),
(\ref{phi_in_domains}) and (\ref{theta_static_in_TDW}), respectively.

Eq. (\ref{finiteTMF_velocity}) shows that the presence of TMFs does improve the TDW propagation velocity.
Firstly, a non-zero $\theta_{\infty}$ leads to
\begin{equation}\label{finiteTMF_vfactor1}
    u(\theta_{\infty})\equiv \frac{2\cos\theta_{\infty}}{2\cos\theta_{\infty}-(\pi-2\theta_{\infty})\sin\theta_{\infty}} >1
\end{equation}
Secondly, from Eqs. (\ref{Walker_phi_belowHW}), (\ref{Walker_velocity_belowHW}), (\ref{phi_in_domains}) and (\ref{theta_static_in_TDW}),
one can always make $\Delta_{\mathrm{Walker}}(H_1)=\Delta(\phi_{\infty})$ by appropriate choice of $\Phi_{\perp}$.
It is worth of investigating the behavior of $u(\theta_{\infty})$ when $\theta_{\infty}\rightarrow \pi/2$
(i.e. $H_{\perp} \rightarrow H_{\perp}^{\mathrm{max}}$).
Let $\pi/2-\theta_{\infty}\equiv \delta$, by noting $\cos\theta_{\infty}=\sin\delta \sim \delta$
and $\sin\theta_{\infty}=\cos\delta\sim (1-\delta^2/2)$, calculation finally yields,
\begin{equation}\label{finiteTMF_vfactor2}
    u(\theta_{\infty})=u(\delta)\sim 2\delta^{-2}\rightarrow +\infty
\end{equation}
which implies that uniform TMFs can greatly enhance the TDW velocity in traveling-wave mode.
Of course, the infinity can not be reached since at that moment the TDW region tends to expand to the entire wire.

In addition, $u(\theta_{\infty})$ obtained here differs from
``$2\cos\theta_{\infty}/[2-(\pi-2\theta_{\infty})\tan\theta_{\infty}]$" appeared in Refs.\cite{JAP_103_073906_2008,Sobolev1,Sobolev2,Sobolev3}
by a factor ``$\cos\theta_{\infty}$".
This difference is the direct manifestation of our handling way of the TDW twisting.

\section{V. Simulation results}
We perform numerical simulations using OOMMF\cite{OOMMF} package to testify our analytics.
In our simulations, the NW is $10\,\mu \mathrm{m}$ long, $5\,\mathrm{nm}$ thick and $150\,\mathrm{nm}$ wide,
which is a common geometry in real experiments.
The following magnetic parameters are adopted:
$M_s=500\,\mathrm{kA/m}$, $J=40\times10^{-12}\,\mathrm{J/m}$ and $K_1=\mu_0 k_1^0 M_s^2/2=200\,\mathrm{kJ/m^3}$.
The crystalline TMA is modeled by setting $K_2=\mu_0 k_2^0 M_s^2/2=80\,\mathrm{kJ/m^3}$
and the Gilbert damping coefficient is chosen to be $\alpha=0.1$ to speed up the simulation.
Throughout the calculation, the NW is spatially discretized into $5\times5\times5\,\mathrm{nm^3}$ cells
to ensure no meshes outside the structure.
In all the figures, $z_0$ denotes the TDW center which is the algebraic average of the centers of each layer (line of cells at a certain $y$) .

\subsection{A. Static profiles}
First we testify the approximate static profile.
The TMF with $H_{\perp}=200\, \mathrm{Oe}$, $\Phi_{\perp}=5\pi/12$ is chosen as an example.
In OOMMF simulations, we first nucleate
an ideal head-to-head N\'{e}el wall at the center of the NW and let it relax to its static profile.
After that, the uniform TMF is exerted onto the whole NW.
The initial wall then begins to evolve and becomes stable in a few nanoseconds.
The magnetization distribution of the final TDW is read out and compared with analytical results.
This is the whole procedure we perform the numerics for static cases.

\begin{figure}[htbp]
\centering
\scalebox{1.1}[1.1]{\includegraphics[angle=0]{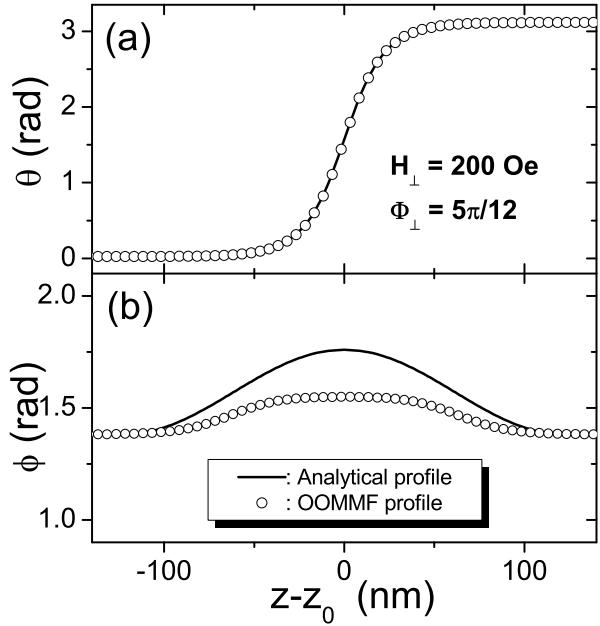}}
\caption{Static profile of TDW in a $5\,\mathrm{nm}\times 150\,\mathrm{nm}\times 10\,\mathrm{\mu m}$ nanowire
under $H_{\perp}=200$ Oe and $\Phi_{\perp}=5\pi/12$ with magnetostatic interaction switched off:
(a) $\theta-$profile; (b) $\phi-$profile.
The solid curves are those from analytical results (\ref{theta_static_in_TDW}) and (\ref{phi_static_in_TDW}), while the open circles
are those from OOMMF simulations. The adopted magnetic parameters are:
$M_s=500\,\mathrm{kA/m}$, $J=40\times10^{-12}\,\mathrm{J/m}$, $K_1=200\,\mathrm{kJ/m^3}$, $K_2=80\,\mathrm{kJ/m^3}$
and $\alpha=0.1$.
}\label{fig2}
\end{figure}

As a first step, the magnetostatic interaction is switched off and the NW becomes a 1D system.
The static $\theta-$ and $\phi-$profiles are translational invariant along the width,
and even independent on the width itself since all the layers are parallel copies.
Simulated static $\theta-$ and $\phi-$profiles of the bottom layer are indicated in Fig. 2a and 2b by open circles.
Meanwhile, analytical results from Eqs.  (\ref{theta_static_in_TDW}) and (\ref{phi_static_in_TDW}) are shown by solid curves.
For $\theta-$profile, analytics and numerics coincide perfectly.
For $\phi-$profile, analytical profile (\ref{phi_static_in_TDW}) perfectly reproduces the continuity and finite twisting region,
while overestimates the twisting amplitude, which confirms our assertion in the end of Sec. III.

\begin{figure}[htbp]
\centering
\scalebox{1.05}[1.05]{\includegraphics[angle=0]{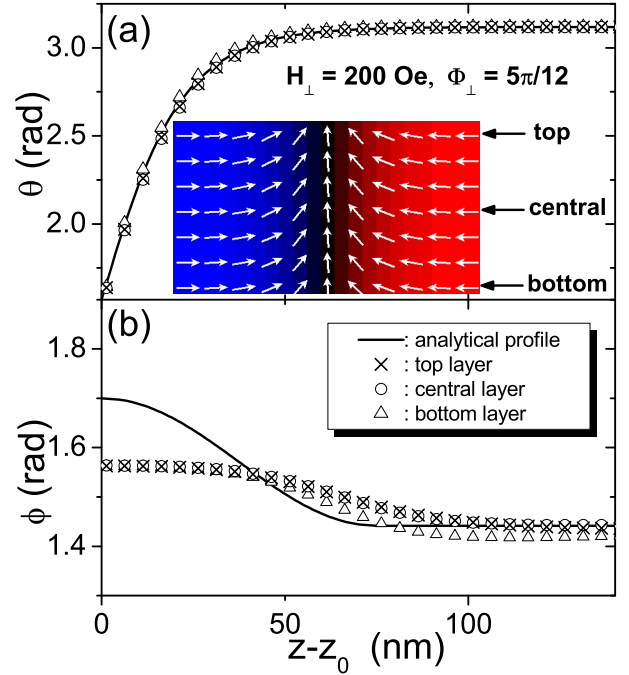}}
\caption{(Color online) Static profile of TDW in a $5\,\mathrm{nm}\times 150\,\mathrm{nm}\times 10\,\mathrm{\mu m}$ nanowire
under $H_{\perp}=200$ Oe and $\Phi_{\perp}=5\pi/12$ with magnetostatic interaction switched on:
(a) $\theta-$profile; (b) $\phi-$profile.
The solid curves are those from analytical results (\ref{theta_static_in_TDW}) and (\ref{phi_static_in_TDW}).
The scattered symbols are those of the bottom (open triangles), central (open circles) and top (crosses) layers
from OOMMF simulations. The magnetic parameters are the same with those in Fig. 2.
Inset of Fig. 3a: screenshot of final TDW in OOMMF simulation. Blue (red) means $m_z>0(<0)$ and black means TDW region.
}\label{fig3}
\end{figure}

In the second step, we switch on the magnetostatic interaction.
The demagnetization factors for this wire geometry are: $D_x=0.94742,\, D_y=0.05185,\, D_z=0.00073$.
They help to give the analytical static $\theta-$ and $\phi-$profiles shown by solid curves in Fig. 3a and 3b.
Interestingly, OOMMF simulations show that even for this width the TDW is still almost 1D,
which can be attributed to the relatively large crystalline anisotropy in wire axis.
This is also true for wires made of real hard-magnetic materials (cobalt, etc.) with easy axis coincident with wire axis.
However, for wires made of soft-magnetic materials (permalloy, etc.), TDWs of this width will generally become two-dimensional (2D).
Discussions for 2D walls would be quite interesting however beyond the scope of this article.
Returning back to our calculation, for each layer, we have calculated the static $\theta-$ and $\phi-$profiles.
For clearance we only show numerical profiles for the bottom, central and
top layers by scattered symbols in fig. 3a and 3b, respectively.
Besides, we always have $\theta(z-z_0)+\theta(z_0-z)\equiv \pi$ and $\phi(z-z_0)\equiv \phi(z_0-z)$  for static cases.
Thus we only show data in the region $z\geq z_0$ in Fig. 3 for clarity.
For polar angle distribution, the three numerical profiles all coincide with the analytical ones very well,
which confirms the quasi-1D nature of the TDW.
For azimuthal angle, again the analytical profile (\ref{phi_static_in_TDW}) overestimates the twisting amplitude.
In addition, the prism geometry will generate extra magnetic charges at the sharp edges of the wire, thus makes the
azimuthal angles of the bottom and top layers slightly different from that of the central layer.

It is also interesting to point out that the $\theta-$profile in Fig. 3a are almost the same with that in Fig. 2a,
which comes from the fact that $(D_y-D_z)\ll k_1^0$ for this wire geometry.
On the contrary, the $\phi-$profiles in Fig. 2b and Fig. 3b differ a lot.
This is the direct consequence of the fact that $(D_x-D_y)$ is comparable with $k_2^0$.
These results validate the feasibility of the ``non-local to local" simplification of magnetostatic interaction 
in thin enough NWs made of hard-magnetic materials.

\subsection{B. Dynamical behaviors}
Next we move to dynamical behaviors. The magnetostatic interaction is always switched on to mimic real experiments.
For magnetic parameters and wire geometry mentioned above,
$k_1 M_s=8.32\,\mathrm{kOe}$, $H_W=\alpha k_2 M_s/2\approx 440\,\mathrm{Oe}$.
All dynamical calculations are constrained within the range of $0\leq H_1\leq 260\,\mathrm{Oe}$ to avoid unnecessary complexity
from the possible collapse of traveling-wave mode under high axial fields.
We choose $H_{\perp}=1\, \mathrm{kOe}$ and $\Phi_{\perp}=\pi/3$ as an example and perform the corresponding analytical
and numerical calculations. The results are shown in Fig. 4.

\begin{figure}[htbp]
\centering
\scalebox{0.75}[0.75]{\includegraphics[angle=0]{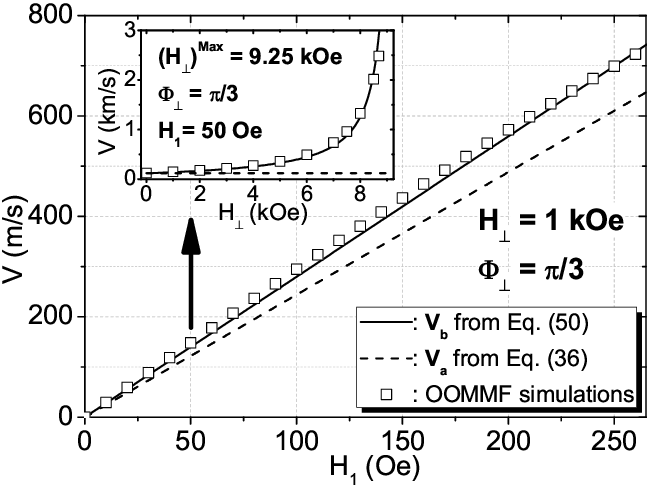}}
\caption{TDW velocity in a $5\,\mathrm{nm}\times 150\,\mathrm{nm}\times 10\,\mathrm{\mu m}$ nanowire
under $H_{\perp}=1\, \mathrm{kOe}$ and $\Phi_{\perp}=\pi/3$ in traveling-wave mode as a function of $H_1$.
The magnetostatic interaction is always switched on.
Solid and dash lines are those from ``finite TMF case" result (\ref{finiteTMF_velocity})
and ``small TMF case" result (\ref{smallTMF_velocity}), respectively.
The open squares are those from OOMMF simulations.
Magnetic parameters are the same with those in Fig. 2.
Inset: TDW velocity enhancement by TMFs with $\Phi_{\perp}=\pi/3$ at $H_1=50\, \mathrm{Oe}$.
}\label{fig4}
\end{figure}

In this figure, the solid line is directly calculated from
the $V_b(H_1)$ relationship (\ref{finiteTMF_velocity}) in ``finite TMF case" section, while the dash line comes from the
$V_a(H_1)$ relationship (\ref{smallTMF_velocity}) in ``small TMF case" section.
Meanwhile, OOMMF velocities are calculated as $\langle\frac{\mathrm{d}z_0}{\mathrm{d}t}\rangle$ and indicated in Fig. 4 by open squares.
It is clear that Eq. (\ref{finiteTMF_velocity}) fits the simulation results very well, which implies that it does capture
the core issues.
To further verify this conclusion, we fix the axial driving field at $H_1=50\, \mathrm{Oe}$ and the TMF orientation angle at $\Phi_{\perp}=\pi/3$,
meanwhile vary the TMF strength in the range $0<H_{\perp}<H_{\perp}^{\mathrm{max}}=9.25\,\mathrm{kOe}$.
The analytical results from Eq. (\ref{finiteTMF_velocity}) and OOMMF simulations are shown by solid curve and open squares
in the inset of Fig. 4, respectively.
One can see that they coincide with each other perfectly, which further confirms the validity of Eq. (\ref{finiteTMF_vfactor2}).
In summary, these results clearly present the extent and boundary of the ``velocity-enhancement" effect of TMFs to TDWs in biaxial NWs.

\begin{figure}[htbp]
\centering
\scalebox{0.9}[0.9]{\includegraphics[angle=0]{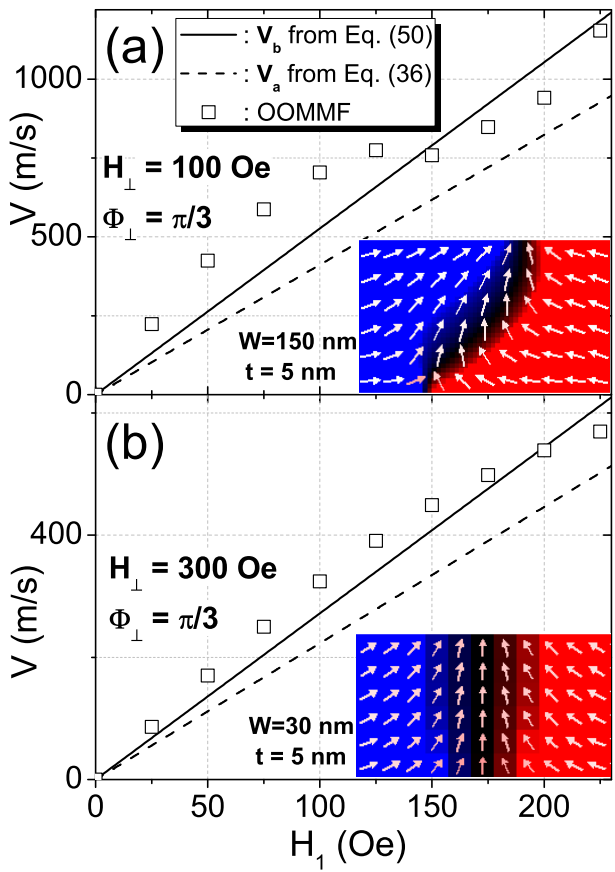}}
\caption{(Color online) TDW velocity in traveling-wave mode as a function of $H_1$ in permalloy nanowires
under $H_{\perp}=1\, \mathrm{kOe}$ and $\Phi_{\perp}=\pi/3$ for two typical wire geometries:
(a) $5\,\mathrm{nm}\times 150\,\mathrm{nm}\times 10\,\mathrm{\mu m}$;
(b) $5\,\mathrm{nm}\times 30\,\mathrm{nm}\times 10\,\mathrm{\mu m}$.
The magnetostatic interaction is always switched on.
Solid and dash lines are those from ``finite TMF case" result (\ref{finiteTMF_velocity})
and ``small TMF case" result (\ref{smallTMF_velocity}), respectively.
The open squares are those from OOMMF simulations.
Magnetic parameters are: $M_s=860\,\mathrm{kA/m}$, $J=13\times10^{-12}\,\mathrm{J/m}$, $K_1=K_2=0$
and $\alpha=0.1$.
The insets are screenshots of TDWs in OOMMF simulations for the corresponding geometry.
Blue (red) means $m_z>0(<0)$ and black means TDW region.
Note that in the inset of Fig. 5a, each arrow represents the average orientation of $25\times 25$ basic calculation nodes.
}\label{fig5}
\end{figure}

In the end of this work, we perform some numerical calculations in permalloy (Py, a typical soft-magnetic material) NWs.
In the first case, the wire geometry is also $5\,\mathrm{nm}\times 150\,\mathrm{nm}\times 10\,\mathrm{\mu m}$.
For Py NWs of this size, TDWs therein have obvious 2D structures when switching on the magnetostatic interaction.
Their dynamical behavior is much more complicated than that in hard-magnetic materials, even in traveling-wave mode.
TDWs will be stretched, distorted and even wiped out as the axial driving field gets large.
In principle, our 1D analytics would not be a good description to this case.
This is confirmed by data in Fig. 5a, in which the open squares (OOMMF simulations) can not be well described
by the solid line coming from Eq. (\ref{finiteTMF_velocity}).
With the thickness unchanged, if we shrink the wire width, the TDW gradually loses 2D details and becomes more and more 1D.
We can expect the coincidence between the 1D analytics and OOMMF simulations should get better.
This is confirmed by data in Fig. 5b, in which similar calculations are performed for a Py NW with
$5\,\mathrm{nm}$ thick and $30\,\mathrm{nm}$ wide.
These results show the limitation of our analytics: it does apply only to 1D (or quasi 1D) TDW systems.

\section{VI. Discussions}
In our analytics, the non-local magnetostatic interaction is imitated by local quadratic terms of magnetization.
In this sense, it has not been fully treated hence our DWs under investigation are indeed 1D and do not show triangular ``V" shape.
However, numerical data in Sec. V show that for NWs made of hard-magnetic materials
and with cross-section up to $5\times 150\,\mathrm{nm^2}$, DWs therein are still almost 1D.
For both static and dynamical cases, simulation data agree with our analytics very well.
This means that for these NWs, our imperfect treatment of magnetostatic interaction does capture the core issues.
In these NWs, due to the inevitable twisting, DWs are neither rigorous Bloch walls nor rigorous N\'{e}el walls.
They are not even traditional $180^{\circ}$ walls in the presence of a uniform TMF.
Based on these reasons, we nominate the walls in our work as ``\emph{transverse} DWs(TDWs)".

Next, from the roadmap of field-driven DW dynamics\cite{xrwfield_AOP,xrwfield_EPL}, the TDW velocity is directly
proportional to the energy dissipation rate of the wire.
From the discussions in Sec. III.C, our approximate static TDW profile (also the zero-order solution) with overestimated twisting has higher
magnetic energy density than the real one.
Hence Eq. (\ref{finiteTMF_velocity}) should be an upper bound (not supremum) of the real TDW velocity in biaxial NWs under finite uniform TMFs.
Better decoupling manner of polar and azimuthal degrees of freedom will lead to better estimation of the supremum.
On the other hand, the ``non-local to local" simplification of magnetostatic interaction in thin enough NWs
tends to ignore most frustrated details of magnetization distribution
thus eventually compensates this overestimation of TDW velocity to a great extent.
In summary, it is this competition between these two aspects that results in the perfect coincidence between analytics and numerics in Fig. 4.

At last, we would like to summarize here the advantages and disadvantages of our approach.
Our approximate static solution reproduces perfectly the polar angle profile in real static TDWs,
meanwhile preserves the continuous twisting in azimuthal angle distribution to a great extent.
Based on it, the TDW velocity shows an explicit dependence on TMFs,
which provides the extent and boundary of their ``velocity-enhancement" effect to TDWs in biaxial NWs.
This can be used to explain existing results or even predict future numerical simulations and experimental measures.
Moreover, our deduction process implies a routine in the asymptotic approach: we can
simplify the first-order operators with the help of zero-order equations.
However, there are still places to be improved.
First, we need to seek for more elegant decoupling manners rather than that shown in Eqs. (\ref{AB_to_f1f2}-\ref{f1f2_eq_0}).
Secondly, after decoupling, to obtain the approximate static
$\phi$-profile, we weakened Eq. (\ref{f1f2_eq_0}) to (\ref{f1f2_weakened}).
This is another source of error.
Thirdly, in ``finite TMF case", we have to assume
that $\theta_0$ and $\phi_0$ had been decoupled so that the first-order operators $\mathbf{R}$ and
$\mathbf{S}$ could be simplified.
Further efforts will be performed to these issues.

\section{Acknowledgement}
This work is supported by the National Natural Science Foundation of China (Grant No. 11374088).

\end{document}